# 3D Geological Modeling and Visualization of Rock Masses Based on Google Earth: A Case Study

Gang Mei, John C.Tipper
Institut für Geowissenschaften – Geologie
Albert-Ludwigs-Universität Freiburg
Freiburg im Breisgau, Germany
{gang.mei, john.tipper}@geologie.uni-freiburg.de

Nengxiong Xu
School of Engineering and Technology
China University of Geosciences (Beijing)
Beijing, China
xunengxiong@yahoo.com.cn

*Abstract*— Google Earth (GE) has become a powerful tool for geological modeling and visualization. An interesting and useful feature of GE, *Google Street View*, can allow the GE users to view geological structure such as layers of rock masses at a field site. In this paper, we introduce a practical solution for building 3D geological models for rock masses based on the data acquired by use with GE. A real study case at Haut-Barr, France is presented to demonstrate our solution. We first locate the position of Haut-Barr in GE, and then determine the shape and scale of the rock masses in the study area, and thirdly acquire the layout of layers of rock masses in the Google Street View, and finally create the approximate 3D geological models by extruding and intersecting. The generated 3D geological models can simply reflect the basic structure of the rock masses at Haut-Barr, and can be used for visualizing the rock bodies interactively.

## I. Introduction

Google Earth (GE) is considered as a virtual globe and geographical information program, which has drawn increasing research interests in geospatial applications and technologies. The recent development and scientific applications of GE were summarized and reviewed in [1, 2].

Keyhole Markup Language (KML) has become a standard for manipulating 3D geospatial data, which can be used in GE and the NASA World Wind. In scientific fields, many people have visualized specific data sets using GE based on KML or developed various applications by adding their own data.

Numerous tools or packages have been designed to convert specific data into the format of KML that can be imported to GE and then visualized, include csv2kml [3], KML generators [4], WKML [5], RCMT [6] and S2K [7]. In [8], Ballagh et al described various KML creation methods and gave a guide for selecting tools to author KML for use with scientific data.

Visualization of data sets in geosciences is one of the most interesting fields that GE is applied in. In [9], Sheppard and Cizek reviewed the principles, benefits and risks of using GE for visualizing landscape. While in [10], the GE is adopted to visualize volcanic gas plumes.

Recently, some applications based on GE are developed in various disciplines: Sun et al [11] have developed a web-based visualization platform for climate research based on GE; Chien and Tan [12] adopted GE as a tool to create 2-D hydrodynamic models; De Paor and Whitmeyer [13] developed KML codes to render geological maps and link associated COLLADA models to represent geological and geographical data.

Besides the applications described above, several other GE based applications also have been developed including ATDD [14], GESO [15], Vis-EROS [16], ARMAP [17] and 4DPlates [18]. All of those applications either use GE to acquire specific kinds of data and then visualize the data in GE or import their own data sets into GE and display them.

An important and interesting feature of GE, *Google Street View*, has been integrated into GE that allows the users to view 3D objects such as buildings, structures and rocks on the street level view. This feature can be accepted to display and acquire detailed geological structure such as the layers of rock masses at a field site without planning a trip to visit the real site.

In this paper, we describe a useful approach for building the 3D geological model for the rock masses located in the selected study area, Haut-Barr, based on the geological data completely acquired from GE especially by taking advantage of the feature of GE, Google Street View.

This paper is organized as follows. In Sect.2, the outline of building the 3D geological model based on GE is summarized. Sect.3 introduces the implementation of a real study case and shows how to build 3D geological models of the selected study area. Finally, in Sect.4 we conclude our work in this paper.

## II. The approach

The 3D geological modeling technique is a powerful tool to depict and visualize geological and engineering realities. The acquisition of various data, e.g., the location, the geometric properties and geologic properties, of geological objects such as rock masses is the fundamental issue of building geological models and visualizing.

Taking a rock mass as an example, we need to determine where it locates, its shape and scale, the layers and faults. In some cases, we maybe should know the age and material of the rock mass. The conventional approach to obtain these data is to make a field trip to the site where the rock mass locates, and then record the needed data by measuring and experimental test.

By using GE, especially taking advantages of the feature, Google Street View, people can virtually visit the target field site and view those rock masses they are interested in.

This research was supported by the Natural Science Foundation of China (Grant Numbers 40602037 and 40872183) and Fundamental Research Funds for the Central Universities of China.

In this paper, we present a simple solution of building 3D geological models without needing to make a time-cost field trip. All necessary data of the rock masses such as the locations, scales and shape, and layers and faults, are completely obtained by use with GE.

Our procedure of constructing the 3D geological models for rock masses can be summarized mainly in five steps:

Step 1: Determine the locations of the rock masses by using GE's location searching simply.

Step 2: Obtain the shapes and scales of the rock masses by measuring on the GE satellite images.

Step 3: View the rock structure such as layers and faults of the rock masses in the Google Street View.

Step 4: Draw the planar wireframe/line models of the rock masses according to their shape, scale, layers and faults.

Step 5: Building the 3D geological models by extruding the planar wireframe models to 3D ones and then conducting the intersection of the 3D models created by extruding.

The key issue in our procedure is to acquire the structure of rock masses. And in this paper, only simple structure such as layers of rock masses needs to be modeled for simulation.

### III. CASE STUDY

#### A. The study area

The Haut-Barr / Château de Hohbarr, as shown in Figures 1 and 3, is a medieval castle, first built in 1100 on a rose-colored sandstone rock 460m above the valley of Zorn and the plane of Alsace in France. The rock bar made of three consecutive rocks: the Septentrional Rock (labeled as 3 in Figure 3), the Median Rock (labeled as 2) and the Southernmost Rock that is best known as Markenfels (labeled as 1), has an altitude of 470m. Two of rock masses that labeled as 1 and 2, are joined by a foot bridge named Devil's Bridge. Since the 12th century, the Haut-Barr Castle has been built by Rodolphe, bishop of Strasbourg.

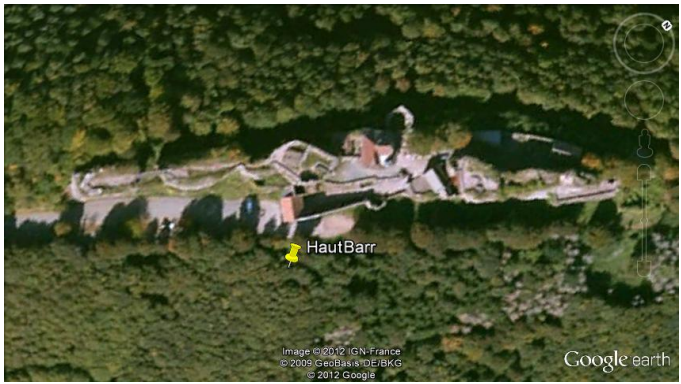

Figure 1. Haut-Barr displayed with GE in the top view

#### B. Implementation of modeling

##### 1) Shape and scale of the rock masses

The shape of Haut-Barr displayed with GE in the top view is shown in Figure 1. We divide the rock masses into four parts: the three separated rock masses that labeled as 1 to 3 in Figure 3 and the base rock bar (labeled as 4) that the above three rock masses lay on.

Figure 2 is the rough and planar wireframe model of Haut-Barr, which is drawn according to the view in Figure 1. In addition, this line model is also composed of four parts.

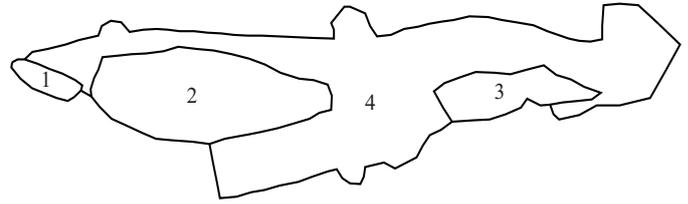

Figure 2. Planar wireframe model of Haut-Barr (in the top view)

Above description is for acquiring the shape. Comparing to obtain the shape of Haut-Barr, the scale is much easier to determine. We first create a minimum bounding box, which virtually encloses the Haut-Barr; and the things we need to do next are to determine the three parameters of the minimum bounding box: length, width and height. Acquisition of the scale of Haut-Barr is exactly to obtain these parameters.

We simply use the measuring toolbar of GE to calculate the maximum length and width of Haut-Barr in meters. And we receive the altitude of the ground surface displaying on the screen by moving the cursor to the right positions on the GE client. The maximum altitude of the rock masses is 470m. The height of the visible rock masses is just the difference of the maximum altitude, 470m, to the altitude of the terrain.

The average altitude of the terrain displayed in GE is 425m, thus the maximum visible length of the rock masses is 45m. We simply extend this distance to be a little larger, i.e. 50m, with considering the underground parts of the rock masses.

In summary, the scale of Haut-Barr, which is represented using a bounding box, consists of 255m, 70m and 50m in length, width and height, respectively.

##### 2) Structures of the rock masses

In engineering geology and geotechnical engineering, the rock structure of rock masses including rock bodies and various structural planes such as layers, faults and joints is the key issue that needs to be considered.

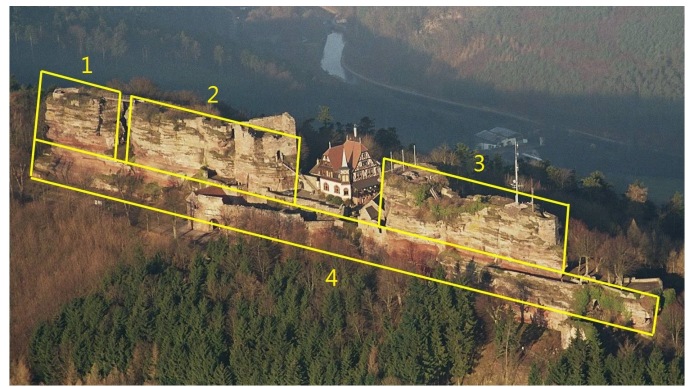

Figure 3. A real view of Haut-Barr. The picture is available at http://www.rotarysaverne.org/un-nouveaux-siege/

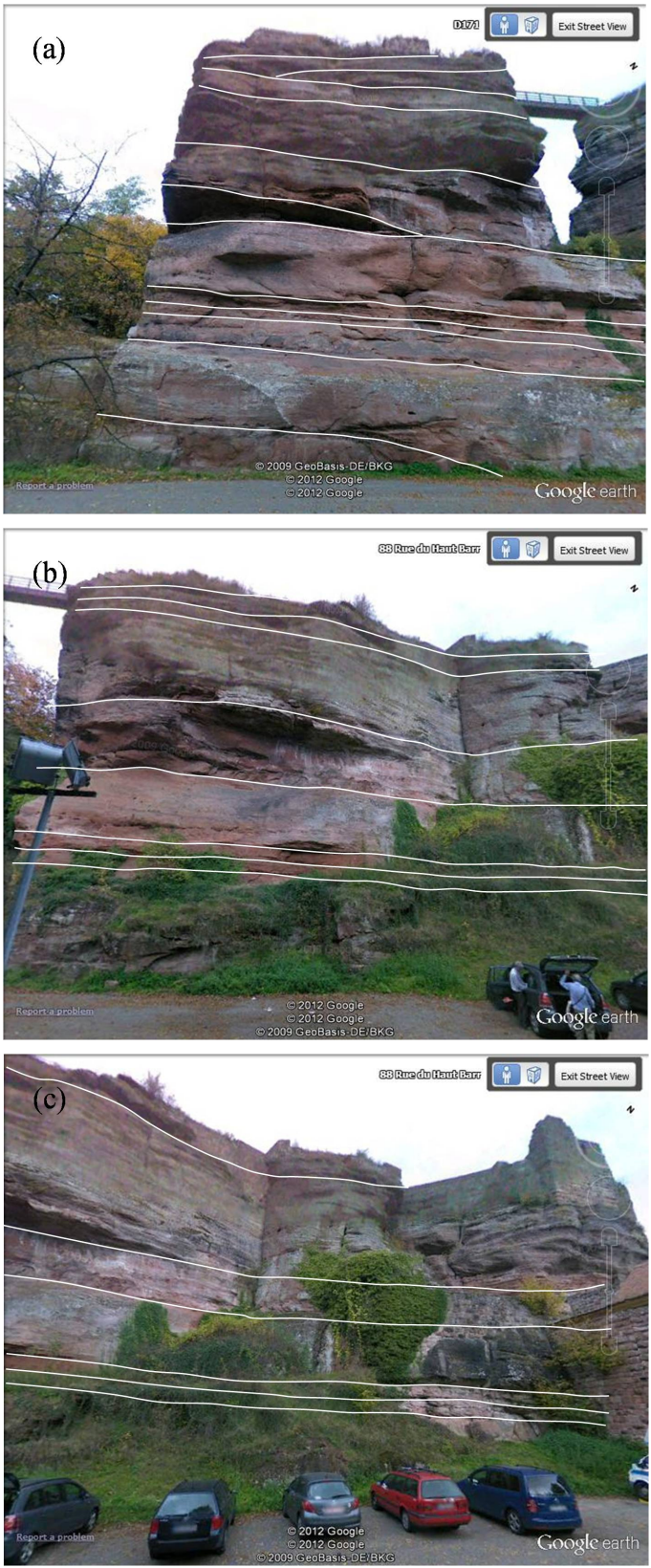

In general, engineers need to make a field trip to the site where the rock masses locate to view and record the details of rock structure. This approach of acquiring the geometric and geologic information about the rock masses is quite meaningful and essential, especially when the rock structure is complex.

Taking advantage of the feature, Google Street View, we visit Haut-Barr in a virtual field trip and view the layers of the rock masses, as shown in Figure 4. According to the layers displayed in GE, a wireframe model that can be approximately represented the layout of layers is created, shown in Figure 5.

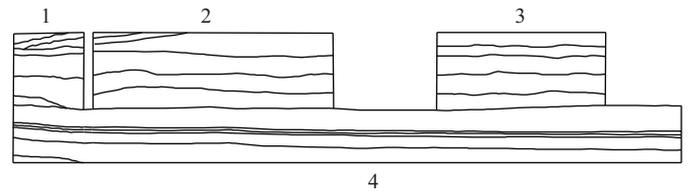

Figure 5.   Layout of the layers at Haut-Barr (viewed along the south side)

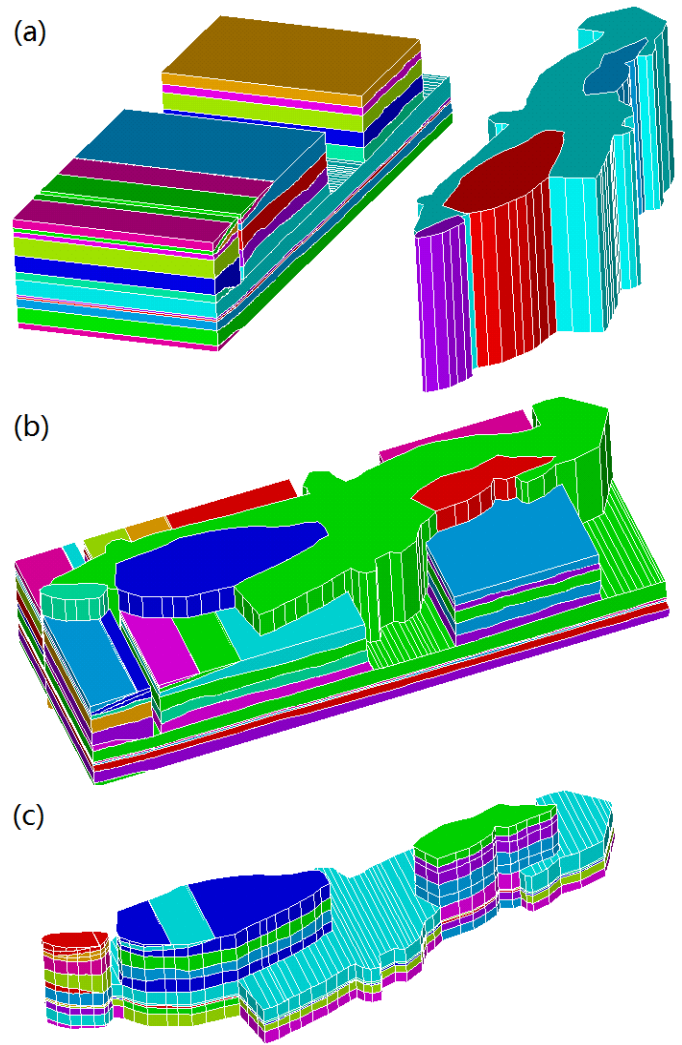

Figure 4.   Layers of Haut-Barr displayed in the Google Street View. (a) Rock mass 1 and part of rock mass 4; (b) Part of rock mass 2 and another part of rock mass 4; (c) The rest part of rock mass 2 and part of rock mass 4

Figure 6.   3D geological models of Haut-Barr. (a) 3D models created by extruding planar wireframe models; (b) Overlapping of the two 3D models formed in (a); (c) Final 3D geological model of Haut-Barr by intersecting

### 3) 3D geological models of the rock masses

The original 3D geological models are created by extruding the wireframe models shown in Figures 2 and 5 along specific directions, as shown in Figure 6a. These models are of 255m, 70m and 50m in length, width and height, respectively.

In Figure 6b, the original 3D models created by extruding are put together to show the overlap. And then the intersection of this pair of 3D models is conducted to generate the final and approximate 3D geological models of the rock masses at Haut-Barr, as shown in Figure 6c.

## IV. SUMMARY

In this paper, we introduce a practical solution for building the 3D geological models for rock masses based on the data sets that are completely acquired by use with GE, especially the interesting and important feature of GE, Google Street View.

The data sets of the rock masses obtained from GE include the location, geometric properties such as shape and scale, and rock structure such as layers. We also provide a real study case to demonstrate the procedure of acquiring desired data by using GE, and then build 3D geological models for Haut-Barr.

We first locate the position of Haut-Barr in GE, and then determine the shape and scale of rock masses in the study area, and thirdly acquire the layout of layers of rock masses by use with the GE feature Google Street View, and finally generate the approximate 3D geological models by extruding the planar wireframe models to form original 3D models and conducting the intersecting of the two original 3D models.

The generated 3D geological models can represent the basic structure of the rock masses at Haut-Barr well, and can be used for visualizing the rock bodies interactively. In the future, these 3D geological models can be modified and edited to be more complex to reflect more realities of the rock masses, and can be used as computational models in numerical simulation such as analyzing the stabilities of rocks.

## ACKNOWLEDGMENT

The corresponding author Gang Mei would like to thank Chun Liu at Universität Kassel for sharing several interesting and valuable ideas.

## REFERENCES

[1] P. Kingsbury and J. P. Jones Iii, "Walter Benjamin's Dionysian Adventures on Google Earth," *Geoforum,* vol. 40, pp. 502-513, 2009.

[2] L. Yu and P. Gong, "Google Earth as a virtual global tool for Earth science applications at the global scale: progress and perspectives," *International Journal of Remote Sensing,* vol. 33, pp. 3966-3986, 2011.

[3] N. H. Oberlies, J. I. Rineer, F. Q. Alali, K. Tawaha, J. O. Falkinham Iii, and W. D. Wheaton, "Mapping of sample collection data: GIS tools for the natural product researcher," *Phytochemistry Letters,* vol. 2, pp. 1-9, 2009.

[4] Y. Yamagishi, H. Yanaka, K. Suzuki, S. Tsuboi, T. Isse, M. Obayashi, H. Tamura, and H. Nagao, "Visualization of geoscience data on Google Earth: Development of a data converter system for seismic tomographic models," *Computers & Geosciences,* vol. 36, pp. 373-382, 2010.

[5] G.-T. Chiang, T. O. H. White, M. T. Dove, C. I. Bovolo, and J. Ewen, "Geo-visualization Fortran library," *Computers & Geosciences,* vol. 37, pp. 65-74, 2011.

[6] L. Postpischl, P. Danecek, A. Morelli, and S. Pondrelli, "Standardization of seismic tomographic models and earthquake focal mechanisms data sets based on web technologies, visualization with keyhole markup language," *Computers & Geosciences,* vol. 37, pp. 47-56, 2011.

[7] T. G. Blenkinsop, "Visualizing structural geology: From Excel to Google Earth," *Computers & Geosciences,* vol. 45, pp. 52-56, 2012.

[8] L. M. Ballagh, B. H. Raup, R. E. Duerr, S. J. S. Khalsa, C. Helm, D. Fowler*, et al.*, "Representing scientific data sets in KML: Methods and challenges," *Computers & Geosciences,* vol. 37, pp. 57-64, 2011.

[9] S. R. J. Sheppard and P. Cizek, "The ethics of Google Earth: Crossing thresholds from spatial data to landscape visualisation," *Journal of Environmental Management,* vol. 90, pp. 2102-2117, 2009.

[10] T. E. Wright, M. Burton, D. M. Pyle, and T. Caltabiano, "Visualising volcanic gas plumes with virtual globes," *Computers & Geosciences,* vol. 35, pp. 1837-1842, 2009.

[11] X. Sun, S. Shen, G. G. Leptoukh, P. Wang, L. Di, and M. Lu, "Development of a Web-based visualization platform for climate research using Google Earth," *Computers & Geosciences,* vol. 47, pp. 160-168, 2012.

[12] N. Q. Chien and S. Keat Tan, "Google Earth as a tool in 2-D hydrodynamic modeling," *Computers & Geosciences,* vol. 37, pp. 38-46, 2011.

[13] D. G. De Paor and S. J. Whitmeyer, "Geological and geophysical modeling on virtual globes using KML, COLLADA, and Javascript," *Computers & Geosciences,* vol. 37, pp. 100-110, 2011.

[14] A. Chen, G. Leptoukh, S. Kempler, C. Lynnes, A. Savtchenko, D. Nadeau, and J. Farley, "Visualization of A-Train vertical profiles using Google Earth," *Computers & Geosciences,* vol. 35, pp. 419-427, 2009.

[15] R. van Lammeren, J. Houtkamp, S. Colijn, M. Hilferink, and A. Bouwman, "Affective appraisal of 3D land use visualization," *Computers, Environment and Urban Systems,* vol. 34, pp. 465-475, 2010.

[16] G. D. Standart, K. R. Stulken, X. Zhang, and Z. L. Zong, "Geospatial visualization of global satellite images with Vis-EROS," *Environmental Modelling & Software,* vol. 26, pp. 980-982, 2011.

[17] G. Walker Johnson, A. G. Gaylord, J. C. Franco, R. P. Cody, J. J. Brady, W. Manley, M. Dover, D. Garcia-Lavigne, R. Score, and C. E. Tweedie, "Development of the Arctic Research Mapping Application (ARMAP): Interoperability challenges and solutions," *Computers & Geosciences,* vol. 37, pp. 1735-1742, 2011.

[18] S. R. Clark, J. Skogseid, V. Stensby, M. A. Smethurst, C. Tarrou, A. M. Bruaset, and A. K. Thurmond, "4DPlates: On the fly visualization of multilayer geoscientific datasets in a plate tectonic environment," *Computers & Geosciences,* vol. 45, pp. 46-51, 2012.